\documentclass[aps,prl,reprint,amsmath,amssymb]{revtex4-2}

\usepackage{graphicx}
\usepackage{bm}
\usepackage{physics}
\usepackage{hyperref}
\newcommand{\AXXZ}{\mathcal{A}^{\mathrm{XXZ}}_{\mathrm{NN}}}
\newcommand{\AHub}{\mathcal{A}^{\mathrm{Hub}}_{\mathrm{NN}}}
\newcommand{\AXXZOne}{\mathcal{A}^{\mathrm{XXZ}}_{\mathrm{NN+1}}}
\newcommand{\AHubOne}{\mathcal{A}^{\mathrm{Hub}}_{\mathrm{NN+1}}}
\newcommand{\AHubd}{\mathcal{A}^{\mathrm{Hub}}_{\mathrm{NN+d}}}
\begin{document}

\title{Connectivity Controls Variational Accessibility in Symmetry-Preserving Quantum Circuits}

\author{Sahinur Reja}
\affiliation{Department of Physics, Jadavpur University, Kolkata 700032, West Bengal, India}


\begin{abstract}
Designing expressive quantum circuits for the variational quantum eigensolver (VQE) is a central challenge in simulating interacting many-body systems. Even symmetry-preserving (SP) circuits need not access the full many-body symmetry sector. We demonstrate this explicitly for the spin-$1/2$ XXZ and Hubbard models, where Hamiltonian-inspired nearest-neighbor (NN) SP circuits generate restricted dynamical Lie algebras (DLAs). While these restricted manifolds can accurately represent the ground state near the free-fermion limits, $J_z=0$ for XXZ and $U=0$ for Hubbard, interactions expose inaccessible many-body directions and severely limit VQE performance even at large circuit depth.
Remarkably, a minimal modification of the generator connectivity can reconstruct the full orthogonal DLA within all relevant symmetry sectors studied in this work. For XXZ model, the addition of {\it only} one SP next-to-NN single-excitation connection triggers this dramatic reconstruction, whereas the Hubbard case requires a SP correlated double-excitation;--showing that the relevant connectivity can extend from the spatial lattice to many-body configuration space. This reconstruction enlarges the tangent space and accessibility of the energy-lowering Hamiltonian direction, leading to near-unit ground-state fidelity in VQE simulations. Our results establish a direct link between generator connectivity and variational accessibility and suggest a broader circuit-design principle: identify minimal symmetry-preserving generators that enlarge the accessible many-body space rather than increasing circuit depth alone. This perspective may have broader relevance for quantum many-body simulation at the interface of condensed-matter physics and quantum computing.  
\end{abstract}

\maketitle
\textit{Introduction.---}
Variational quantum algorithms provide a natural framework for addressing
quantum many-body problems on present and near-term quantum processors
\cite{Peruzzo2014,Cerezo2021}.  In particular, the variational quantum
eigensolver (VQE) replaces the exponentially demanding representation of a
many-body wave function by a parametrized quantum circuit whose parameters
are optimized to minimize the energy.  Its practical success, however,
depends critically on the choice of the variational ansatz.  A useful circuit
must be sufficiently expressive to contain the physically relevant states
while remaining shallow, trainable, and compatible with the symmetries and
connectivity of the underlying problem.

Problem-inspired and Hamiltonian variational ans\"atze (HVAs) provide an
appealing route toward this balance by constructing the circuit from
operators related to the target Hamiltonian
\cite{Wecker2015,Wiersema2020,Cerezo2021}.  Such circuits can exhibit
favorable optimization properties compared with generic highly expressive
ans\"atze \cite{Wiersema2020,ParkKilloran2024}, while explicitly preserving
physical symmetries can substantially reduce the variational search space
\cite{Gard2020}.  Nevertheless, symmetry preservation by itself does not
guarantee that a circuit can efficiently explore the relevant symmetry
sector.  A Hamiltonian-inspired ansatz may inherit additional algebraic
constraints from its generator set and connectivity, leaving physically
important directions inaccessible even when the circuit depth is increased.

Recent developments have made this restriction increasingly precise.
The geometry of parametrized quantum circuits can be characterized through
the quantum geometric tensor and its effective dimension, which quantify
the number of locally accessible directions in state space
\cite{Haug2021}. Complementarily, the dynamical Lie algebra (DLA) generated by the circuit operators characterizes its global algebraic accessibility and provides bounds on the rank of the quantum Fisher information matrix~\cite{Larocca2022,Larocca2023}. These results establish a direct connection between circuit generators, controllability, trainability, and variational capacity. More recently, DLAs have also been explored as tools for engineering generator sets and accessible quantum dynamics~\cite{liang2026dla}. Moreover, recent classifications of DLAs for local spin systems have demonstrated that the topology on which the generators act can fundamentally alter the resulting Lie algebra~\cite{Wiersema2024}. Together, these observations suggest that circuit connectivity itself can constitute a fundamental variational resource, rather than merely a hardware constraint.

In this Letter, using two paradigmatic many-body interacting models, the spin-1/2 XXZ and Hubbard model, we explicitly show that Hamiltonian-inspired symmetry-preserving (SP) nearest-neighbor (NN) circuits can generate highly restricted DLA that can't be enlarged by increasing circuit depth ($N_L$) and severely limits the VQE performance away from non-interacting limit ($J_z=0$ for XXZ and $U=0$ for Hubbard model). Remarkably, introducing a single appropriate SP generator ({\it only} a next-to-NN single-excitation connection for XXZ and a correlated double-excitation generator for Hubbard) can reconstruct the full orthogonal DLA within the relevant symmetry sector in all cases studied.

This algebraic reconstruction has a direct manifestation in finite-depth VQE simulation. For both the models, the additional generators sharply enlarge the tangent space and improves its accessibility to energy-lowering Hamiltonian direction, resulting in substantially reduced energy errors and near-unit ground-state fidelity. Our results connect generator structure, variational geometry, and VQE performance, and suggest a route toward designing expressive many-body circuits through minimal structural modifications in the quantum circuits rather than increasing circuit depth alone.

\textit{Symmetry-preserving variational circuits.---}
To test this idea, we consider two paradigmatic interacting many-body systems: the spin-$1/2$ XXZ and the Hubbard models with periodic boundary condition (PBC).  The XXZ\cite{Bethe1931,Baxter1982} Hamiltonian is
\begin{equation}
H_{\rm XXZ}
=
\sum_{i}
\left(
S_i^xS_{i+1}^x+S_i^yS_{i+1}^y
\right)
+
J_z\sum_{i}S_i^zS_{i+1}^z ,
\label{eq:xxz}
\end{equation}
Since $[H_{\rm XXZ},S_{\rm tot}^z]=0$, the dynamics decomposes into sectors of fixed magnetization, equivalently fixed excitation number $N_{\uparrow}$. Also, well-known Hubbard model\cite{Hubbard1963,LiebWu1968} reads
\begin{equation}
H_{\rm Hub}
=
-t\sum_{i,\sigma}
\left(
c_{i\sigma}^{\dagger}c_{i+1\sigma}
+
{\rm H.c.}
\right)
+
U\sum_i n_{i\uparrow}n_{i\downarrow},
\label{eq:hubbard}
\end{equation}
for which $[H_{\rm Hub},N_{\uparrow}]=[H_{\rm Hub},N_{\downarrow}]=0.$


\begin{figure}[t]
	\centering
	   \includegraphics[width=0.9\columnwidth]{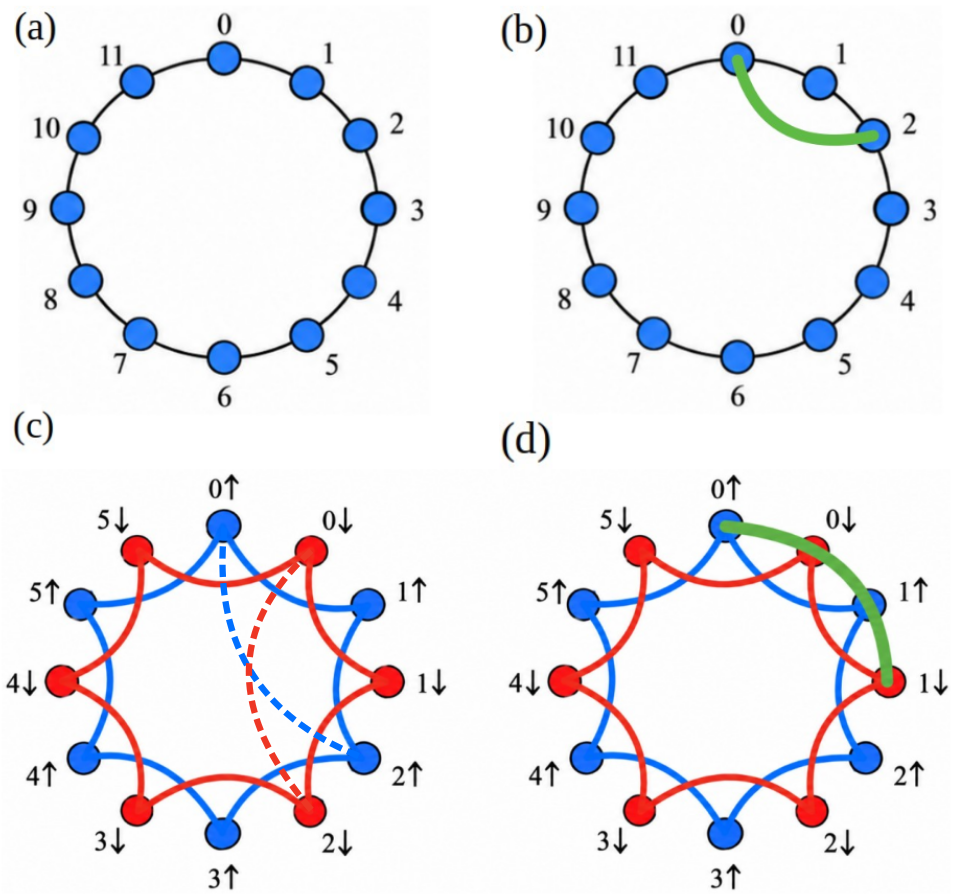}
	\caption{
	Twelve-qubit schematics of various ans\"atze:
	(a) $\AXXZ$:--XXZ ring with two-qubit single-excitation (SE) connections on NN sites only. (b) $\AXXZOne$:-- $\AXXZ$ plus {\it only} one SE connection (green bond) on next-to-NN site. (c)$\AHub$ represents Hubbard ring with NN site hopping excitations (solid blue and red bonds for $\uparrow$ and $\downarrow$-spins respectively) whereas addition of {\it one} next-to-NN connection (blue and red dashed lines for $\uparrow,\downarrow$-spins) gives $\AHubOne$ (d) $\AHubd$:--$\AHub$ plus {\it only} one NN sites four-qubit double-excitation connection (green bond).}
	\label{fig:ansatz_circut}
\end{figure}

To target a given symmetry sector, we initialize the circuit with an occupation-number basis state belonging to that sector. For both the XXZ and Hubbard models, the variational NN circuit is then constructed from symmetry-preserving two-qubit single-excitation (SE) gates,
which generates a rotation in the subspace {$|01\rangle,|10\rangle$}, corresponding to the coherent exchange of $\uparrow$ and $\downarrow$-spin between the two sites while preserving the total magnetization $S_z^{\rm tot}$. Figure.~\ref{fig:ansatz_circut}(a) represents the qunatum circuit with SE gates applied to all NN qubit pairs. We name this simple Hamiltonian-inspired ansatz as $\AXXZ$. We compare its performance with a minimally modified ansatz, namely $\AXXZOne$ in which {\it only} a single genuine next-to-NN (NNN) SE connection (green bond) has been added as depicted in Fig~\ref{fig:ansatz_circut}(b).

For the Hubbard ansatz, namely $\AHub$ as depicted in Fig.~\ref{fig:ansatz_circut}(c), the same NN SE gates are applied separately to the spin-$\uparrow$ and spin-$\downarrow$ orbitals, corresponding to spin-conserving electron hopping (blue and red bonds, respectively). To isolate the role of circuit connectivity, we compare $\AHub$ with two minimally modified SP circuit architectures. The first ansatz $\AHubOne$ is obtained by adding a single genuine NNN SE connection (blue, red dashed lines) to $\AHubOne$ circuit. For the second one, namely $\AHubd$, we introduce {\it only} one SP four-qubit double-excitation (DE) gate in $\AHub$ that coherently rotate between the configurations $|0011\rangle$ and $|1100\rangle$. Depending on the orbital arrangement, this operation describes two distinct processes. A \emph{doublon-transfer} excitation moves an $\uparrow\downarrow$ pair from one site to another, coherently coupling configurations with an empty and a doubly occupied site. An \emph{exchange-type} excitation instead simultaneously transfers $\uparrow$ and $\downarrow$ electrons in opposite directions between two sites, thereby exchanging their spin occupations. Both processes preserve the given $S_z^{tot}$ while introducing correlated directions that cannot be generated by independent SE connections alone. See supplemental materials for more details.

\begin{figure*}[t]
    \centering
    \includegraphics[width=0.99\textwidth]{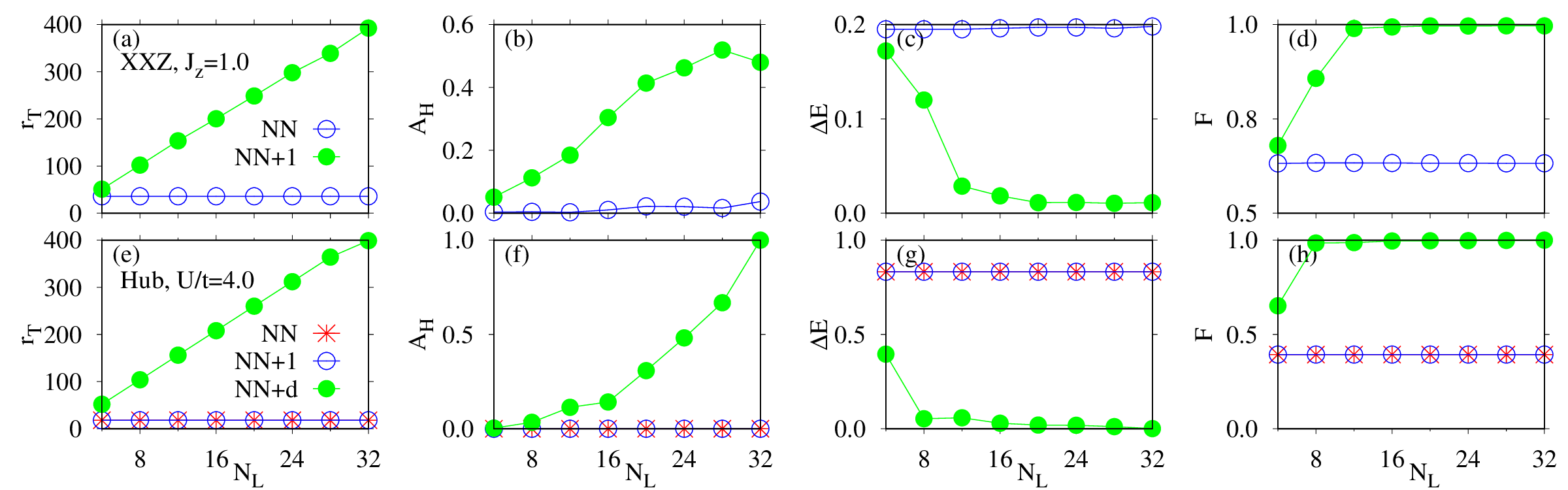}
    \caption{Circuit depth ($N_L$) dependence of variational accessibility and VQE performance of various ans\"atze. For the XXZ ring (top row) with $N=12, S_z^{tot}=0$ and $J_z=1$: (a) Tangent-space rank ($r_T$), (b) Hamiltonian alignment $A_H$, (c) energy error
    $\Delta E=|E_{\rm VQE}-E_0|$, and (d) ground-state fidelity
    $F=|\langle\psi_0|\psi(\boldsymbol{\theta})\rangle|^2$  are shown against $N_L$ for ans\"atze $\AXXZ$ and $\AXXZOne$, indicated by empty (blue) and filled (red) circles, respectively. The corresponding results for $12-$qubit Hubbard ring with $N=6, N_\uparrow=N_\downarrow=3$ and $U/t=4$ are shown in (e)--(h) for ans\"atze $\AHub, \AHubOne$ and $\AHubd$ by empty (blue) circle, star (red) point and filled (green) circles, respectively.}
    \label{fig:vqe_results}
\end{figure*}
\textit{Connectivity-induced Lie-algebra reconstruction.---}
To characterize the global accessibility of the circuit, we consider the
dynamical Lie algebra (DLA)\cite{Larocca2022,DAlessandro2007}
generated by repeated commutators and real linear combinations of the
anti-Hermitian circuit generators.

For XXZ model with $N$ sites and $N_{\uparrow}=k$ excitations, the Hilbert space dimension of the corresponding symmetry sector is $D_{N,k}={^N}C_k$. For the $\AXXZ$ circuit, the generators under the Jordan--Wigner transformation take the form: $L_{ij}=c_i^{\dagger}c_j-c_j^{\dagger}c_i$ which satisfy $[L_{ij},L_{kl}]=\delta_{jk}L_{il}-\delta_{ik}L_{jl}-\delta_{jl}L_{ik}+\delta_{il}L_{jk}$. Thus, the connected NN generator set realizes a special orthogonal $\mathfrak{so}(N)$ algebra with Lie closure dimension $d_{Lie}(\AXXZ)=N(N-1)/2$. Although the symmetry-sector dimension $D_{N,k}$ grows combinatorially with $N$, the NN-generated Lie algebra dimension grows only quadratically.

A genuine NNN generator has a qualitatively different
Jordan--Wigner structure.  For example, the connection $(0,2)$ in $\AXXZOne$ contains
the intermediate parity operator,
\begin{equation}
L_{02}
\longleftrightarrow
c_0^{\dagger}Z_1c_2
-
c_2^{\dagger}Z_1c_0 ,
\qquad
Z_1=1-2n_1 ,
\label{eq:extra_JW}
\end{equation}
and hence introduces occupation-conditioned terms outside the quadratic
algebra generated by the $\AXXZ$ circuit.

We calculate the complete Lie closure numerically in each fixed-$k$ sector. For all finite systems and sectors examined, adding a single NNN connection enlarges the DLA to the full special orthogonal Lie algebra $\mathfrak{so}(D_{N,k})$ within the fixed-excitation sector, with $d_{\rm Lie}(\AXXZOne)=D_{N,k}(D_{N,k}-1)/2$. The contrast with the restricted NN circuit $\AXXZ$ becomes particularly pronounced near half filling; for example, for the $(N=8,S_z^{\rm tot}=0)$ XXZ ring, the DLA dimension increases from $28$ to $2415$.

Similarly for the Hubbard model, we analyze the DLA within the fixed $(N_{\uparrow},N_{\downarrow})$ sector. For example, the $8$-qubit Hubbard ring ($N=4$, $N_{\uparrow}=N_{\downarrow}=2$) has Hilbert space dimension $D=36$ and $\dim \mathfrak{so}(D)=630$. The SE ans\"atze remain strongly restricted, with $\{d_{\rm Lie}(\AHub),d_{\rm Lie}(\AHubOne)\}=\{12,30\}$, whereas adding a single DE generator as in $\AHubd$ reconstructs the full algebra, $d_{\rm Lie}=630$ (see supplemental material for DLA results for $S_z^{tot}\ne0$ sectors).

\textit{Variational Accessibility and VQE Performance.---}
The DLA characterizes global algebraic accessibility, whereas an actual
finite-depth VQE circuit explores only the local directions available at
its current variational state.  For a VQE state
$|\psi(\boldsymbol{\theta})\rangle$, the projected tangent
vectors and their Gram matrix can be defined:
\begin{equation}
|t_{\mu}\rangle
=
\left(
1-|\psi\rangle\langle\psi|
\right)
\frac{\partial|\psi(\boldsymbol{\theta})\rangle}
{\partial\theta_{\mu}} ; \,\,
G_{\mu\nu}
=
\mathrm{Re}
\langle t_{\mu}|t_{\nu}\rangle ,
\label{eq:tangent}
\end{equation}

which determines the number of independent locally accessible directions.
We characterize this geometry through the tangent-space rank (TSR) $r_T$.

To determine whether these directions are physically useful rather than
merely numerous, we further measure the alignment of VQE state  $\psi(\boldsymbol{\theta})$ 
with the local energy-lowering direction quantified by the Hamiltonian
accessibility
\begin{equation}
A_H =
\frac{
\left\| P_{\mathcal{T}} (H-E)|\psi\rangle \right\|^2
}{
\left\| (H-E)|\psi\rangle \right\|^2
},
\qquad
E=\langle\psi|H|\psi\rangle,
\label{eq:AH}
\end{equation}
where $P_{\mathcal{T}}$ denotes the projector onto the variational tangent
space $\mathcal{T}$. Thus, $0\leq A_H\leq 1$, with $A_H=1$ indicating that
the instantaneous energy-lowering direction lies entirely within the
accessible tangent space, while $A_H=0$ indicates that it is locally
inaccessible to the ansatz.

All VQE calculations were performed using PennyLane with state-vector simulation and the Adam optimizer. For a given system size and Hamiltonian, all ans\"atze were optimized using identical learning rates, iteration counts, and parameter-initialization protocols. Multiple independent VQE runs were performed using different random initializations of the variational parameters $\boldsymbol{\theta}$, while keeping the same initial basis state within the target symmetry sector for all ans\"atze and runs. The reported quantities are averaged over these independent runs. Exact/Lanczos diagonalization within the corresponding symmetry sector provided the reference ground-state energy $E_0$ and wave function $|\psi_0\rangle$.

Figure~\ref{fig:vqe_results} summarizes the VQE performance of various ans\"atze considered above against circuit depth ($N_L$). 
For the XXZ model with $N=12, S^{tot}_z=0$ at Heisenberg point \(J_z=1\), Fig.~\ref{fig:vqe_results}(a) shows that the TSR $r_T$ for $\AXXZ$ (empty, blue circles) remains strongly restricted to a small fixed value as $N_L$ is increased. This limitation is directly reflected in the VQE performance: the accessibility of the energy-lowering direction 
\(A_H\), shown in Fig.~\ref{fig:vqe_results}(b), likewise remains very low. So, the energy error ($\Delta E$) saturates at a finite value, while the ground-state fidelity remains appreciably below unity even for sufficiently large $N_L$ as depicted in Fig.~\ref{fig:vqe_results}(c) and (d) respectively. Thus, simply repeating the same NN SE gate structure does not remove the underlying algebraic constraint identified by DLA analysis above.

A qualitatively different behavior emerges upon introducing {\it only} one NNN SE connection to $\AXXZ$ [Fig.~\ref{fig:ansatz_circut}(b)]. The resulting circuit, $\AXXZOne$, progressively enlarges the tangent-space rank $r_T$, accompanied by substantial increase in $A_H$ with circuit depth $N_L$. This geometric reconstruction directly translates into improved VQE performance, strongly suppressing $\Delta E$ and driving the ground-state fidelity toward unity. Importantly, circuit depth $N_L\sim 16$ already gives these improvements. This demonstrates that, for the SP XXZ circuit, generator connectivity---rather than parameter count or circuit depth alone---governs access to the relevant many-body directions.

The Hubbard model exhibits the same underlying restriction despite its different Hamiltonian and symmetry structure. At half-filling ($N=6, N_\uparrow=N_\downarrow=3$) and intermediate correlated regime $U/t=4.0$, the tangent-space rank of $\AHub$ remains restricted to $r_T=18$ with increasing $N_L$, accompanied by strongly suppressed accessibility of the energy-lowering direction $A_H$ [Figs.~\ref{fig:vqe_results}(e) and (f)]. This saturation has a simple geometric origin: each $\sigma$-spin sector spans a real Grassmann manifold of dimension $N_\sigma(N-N_\sigma)=9$, giving $r_T=2\times9=18$ for the two independent spin sectors. Consequently, the energy error $\Delta E$ saturates at a finite large value and the ground-state fidelity remains low [Figs.~\ref{fig:vqe_results}(g) and (h)]. In contrast to the XXZ case, adding a single next-to-NN SE gate in $\AHubOne$ provides essentially no improvement, consistent with the restricted DLA found for SE generators.

\begin{figure}[t]
	\centering
	   \includegraphics[width=\columnwidth]{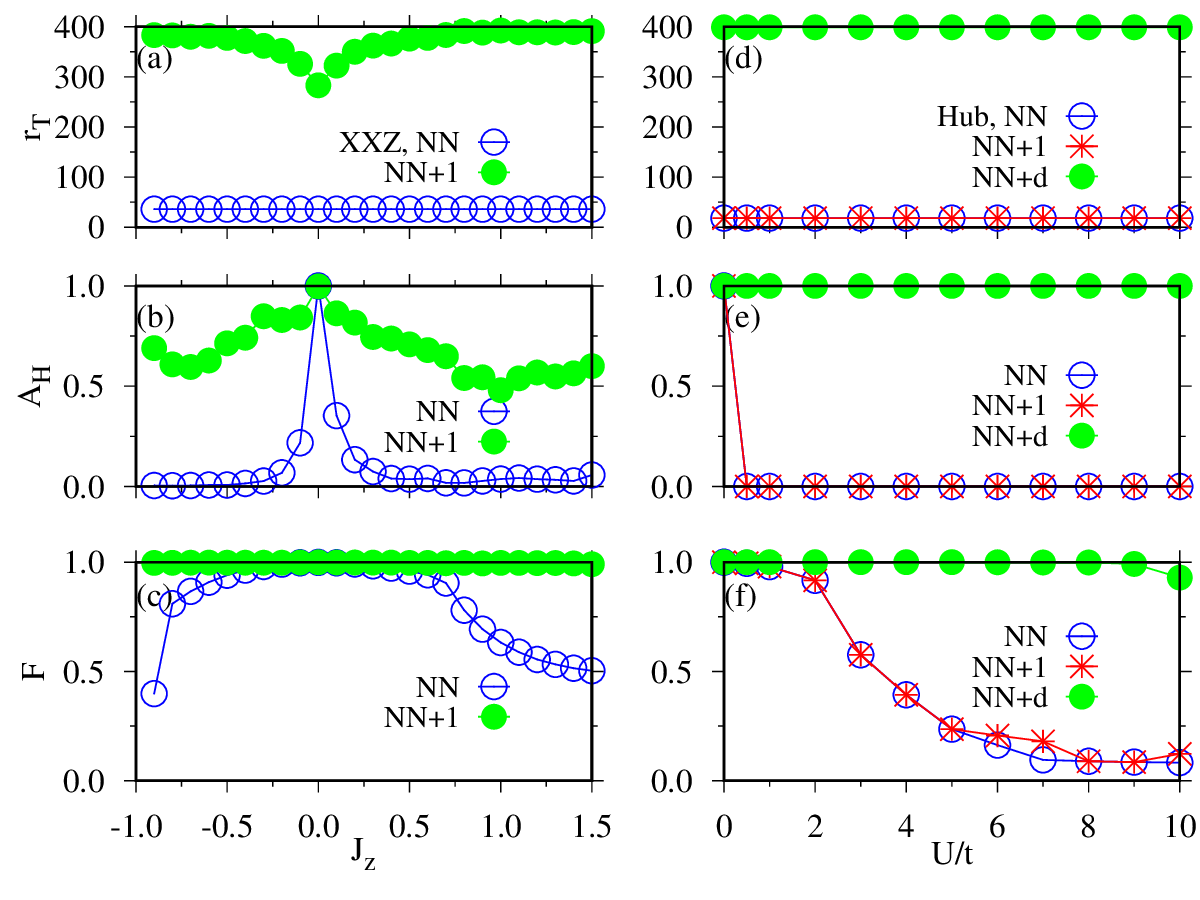}
	\caption{VQE results with $N_L=32$: (a)-(c) depict tangent space rank $r_T$, Hamiltonian alignment $A_H$ and ground-state fidelity $F$   respectively against $J_z$ for spin-1/2 XXZ ring with $N=12, S^{tot}_z=0$. Empty (blue) and filled (green) circles are for ans\"atze $\AXXZ$ and $\AXXZOne$. (d)-(f) show the corresponding results against $U/t$ for $N=6, N_\uparrow=N_\downarrow=3$ Hubbard ring obtained for anzatz $\AHub$ (empty circle, blue), $\AHubOne$ (star point, red)  and $\AHubd$ (filled circle, green) respectively.
	}
	\label{fig:Jz_U_variation}
\end{figure}

This algebraic limitation is lifted when a single SP double-excitation (DE) gate is introduced as in ansatz $\AHubd$. The correlated generator takes the circuit beyond the independent single-particle manifolds, increasing $r_T$ progressively beyond the restricted value and substantially enhancing $A_H$ as circuit depth is increased. As a result, $\Delta E$ decreases sharply and the ground-state fidelity approaches unity (see Fig.~\ref{fig:Jz_U_variation}(e)-(h)). Together with the DLA reconstruction obtained for $\AHubd$, this shows that the essential additional connectivity in the Hubbard model is not simply a longer-range single-particle link, but a connection between correlated many-body configurations.

\textit{Interaction dependence of variational accessibility.---}
Having established the contrasting variational accessibility of the circuits at representative interacting points, $J_z=1$ for XXZ and $U/t=4$ for Hubbard, we now examine how this behavior evolves with interaction strength. We run the simulation on same twelve-qubit rings with $N_L=32$ by varying $J_z$ in the XXZ model and $U/t$ in the Hubbard model for corresponding ans\"atze. Figure~\ref{fig:Jz_U_variation} shows variation of $r_T$, $A_H, F$, for the XXZ (left column) and Hubbard (right column) . A common behavior emerges near the non-interacting limits. Around $J_z=0$, where the XXZ model reduces to the free-fermion XX limit, the NN circuit $\AXXZ$ already provides good accessibility and high $A_H$ and fidelity $F$. Away from this limit, interactions increasingly expose the restricted variational space of $\AXXZ$: the tangent-space rank $r_T$ remains limited, $A_H$ becomes strongly suppressed, and the resulting fidelity deviates substantially from unity. In contrast, the additional NNN connection in $\AXXZOne$ enlarges the accessible tangent space and maintains large $A_H$ and near-unit fidelity across the interacting regime as hsown Fig.~\ref{fig:Jz_U_variation}(a)-(c).

The Hubbard model exhibits the analogous behavior as shown in Fig.~\ref{fig:Jz_U_variation}(d)-(f). At $U=0$, the non-interacting ground state is well captured by the $\AHub$ and $\AHubOne$ circuits, while its variational accessibility (e.g, $A_H$) dratically drops to zero  and fidelity deteriorate as $U/t$ increases. One DE gate in $\AHubd$ supplies the additional correlated directions required at finite interaction, leading to  near-unit Hamiltonian alignment $A_H$ and fidelity $F$. Thus, in both models the restricted circuits can be sufficient near the free-fermion limit, while interactions reveal the need for additional generator connectivity in the many-body variational space. Furthermore, we confirm (see supplemental material) that the underlying connectivity-induced variational accessibility mechanism applies to $S_z^{tot}\ne 0$ sector of XXZ and Hubbard models. 

\textit{Conclusions.---}
We show that symmetry preservation alone does not guarantee sufficient variational accessibility for VQE simulations of interacting many-body systems. In both the spin-1/2 XXZ and Hubbard models, the symmetry-preserving (SP) circuits with nearest-neighbor (NN) single-excitation (SE) connections are confined to restricted dynamical Lie algebras (DLAs), a limitation that cannot be removed by increasing circuit depth. Although these restricted manifolds are sufficient near the free-fermion limits, interactions expose inaccessible many-body directions and strongly degrade VQE performance. We find that minimal modifications of the generator connectivity can dramatically lift these restrictions and restore access to the relevant many-body directions. For XXZ, {\it only} one next-to-NN single-excitation connection is sufficient to reconstruct the full orthogonal DLA within the symmetry sectors studied and recover near-unit ground-state fidelity. For Hubbard model, extending SE connectivity alone is insufficient; instead, a correlated double-excitation is required to achieve the corresponding reconstruction and high-fidelity ground states.

These results identify generator connectivity as a key structural control of variational accessibility. The accompanying enlargement of the tangent space and enhanced accessibility of the energy-lowering Hamiltonian direction provide a finite-depth manifestation of the underlying DLA reconstruction and connect circuit structure directly to VQE performance. Importantly, the Hubbard results show that the relevant notion of connectivity extends beyond the spatial lattice to connections between correlated many-body configurations. This suggests a practical design principle: rather than increasing the depth of a restricted ansatz, identify the minimal SP generators that open the physically relevant many-body directions. Such a perspective may provide a common framework linking many-body correlations, Lie-algebraic structure, and variational circuit design, with broader implications for condensed-matter physics and quantum computing.

\textit{Acknowledgments---}
The author thanks Soumyadeep Sahoo for helpful discussions on general aspects of quantum computing, and Satyabrata Jana and Asim Kumar Ghosh for stimulating discussions. Part of the computation was performed at the Physics Department, Jadavpur University Computer Facility (Sanction No. SR/FST/PS-1/2022/219(C)). 

\bibliography{references}

@article{Peruzzo2014,
	author = {Peruzzo, Alberto and McClean, Jarrod and Shadbolt, Peter and Yung, Man-Hong and Zhou, Xiao-Qi and Love, Peter J. and Aspuru-Guzik, Al{\'a}n and O'Brien, Jeremy L.},
	title = {A variational eigenvalue solver on a photonic quantum processor},
	journal = {Nature Communications},
	volume = {5},
	pages = {4213},
	year = {2014},
	doi = {10.1038/ncomms5213}
}

@article{Cerezo2021,
	author = {Cerezo, M. and Arrasmith, Andrew and Babbush, Ryan and Benjamin, Simon C. and Endo, Suguru and Fujii, Keisuke and McClean, Jarrod R. and Mitarai, Kosuke and Yuan, Xiao and Cincio, Lukasz and Coles, Patrick J.},
	title = {Variational quantum algorithms},
	journal = {Nature Reviews Physics},
	volume = {3},
	pages = {625--644},
	year = {2021},
	doi = {10.1038/s42254-021-00348-9}
}

@article{Wecker2015,
	author = {Wecker, Dave and Hastings, Matthew B. and Troyer, Matthias},
	title = {Progress towards practical quantum variational algorithms},
	journal = {Physical Review A},
	volume = {92},
	pages = {042303},
	year = {2015},
	doi = {10.1103/PhysRevA.92.042303}
}

@article{ParkKilloran2024, 
    author = {Park, C.-Y. and Killoran, N.}, 
    title = {Hamiltonian variational ansatz without barren plateaus}, 
    journal = {Quantum}, 
    volume = {8}, 
    pages = {1239}, 
    year = {2024}, 
    doi = {10.22331/q-2024-02-01-1239} 
}

@article{Wiersema2020,
	author = {Wiersema, Roeland and Zhou, Cunlu and de Sereville, Yvette and Carrasquilla, Juan Felipe and Kim, Yong Baek and Yuen, Henry},
	title = {Exploring Entanglement and Optimization within the Hamiltonian Variational Ansatz},
	journal = {PRX Quantum},
	volume = {1},
	pages = {020319},
	year = {2020},
	doi = {10.1103/PRXQuantum.1.020319}
}

@article{Gard2020,
	author = {Gard, Bryan T. and Zhu, Linghua and Barron, George S. and Mayhall, Nicholas J. and Economou, Sophia E. and Barnes, Edwin},
	title = {Efficient symmetry-preserving state preparation circuits for the variational quantum eigensolver algorithm},
	journal = {npj Quantum Information},
	volume = {6},
	pages = {10},
	year = {2020},
	doi = {10.1038/s41534-019-0240-1}
}

@article{Haug2021,
	author = {Haug, Tobias and Bharti, Kishor and Kim, M. S.},
	title = {Capacity and Quantum Geometry of Parametrized Quantum Circuits},
	journal = {PRX Quantum},
	volume = {2},
	pages = {040309},
	year = {2021},
	doi = {10.1103/PRXQuantum.2.040309}
}

@article{Larocca2022,
	author = {Larocca, Martin and Czarnik, Piotr and Sharma, Kunal and Muraleedharan, Gopikrishnan and Coles, Patrick J. and Cerezo, M.},
	title = {Diagnosing Barren Plateaus with Tools from Quantum Optimal Control},
	journal = {Quantum},
	volume = {6},
	pages = {824},
	year = {2022},
	doi = {10.22331/q-2022-09-29-824}
}

@article{Larocca2023,
	author = {Larocca, Mart{\'i}n and Ju, Nathan and Garc{\'i}a-Mart{\'i}n, Diego and Coles, Patrick J. and Cerezo, Marco},
	title = {Theory of overparametrization in quantum neural networks},
	journal = {Nature Computational Science},
	volume = {3},
	pages = {542--551},
	year = {2023},
	doi = {10.1038/s43588-023-00467-6}
}

@article{Wiersema2024,
	author = {Wiersema, Roeland and K{\"o}kc{\"u}, Efekan and Kemper, Alexander F. and Bakalov, Bojko N.},
	title = {Classification of dynamical Lie algebras of 2-local spin systems on linear, circular and fully connected topologies},
	journal = {npj Quantum Information},
	volume = {10},
	pages = {110},
	year = {2024},
	doi = {10.1038/s41534-024-00900-2}
}

@book{DAlessandro2007,
  author    = {D'Alessandro, Domenico},
  title     = {Introduction to Quantum Control and Dynamics},
  publisher = {Chapman \& Hall/CRC},
  year      = {2007}
}

@misc{liang2026dla,
  author        = {Yanying Liang and Ruibin Xu and Mao-Sheng Li and Haozhen Situ and Zhu-Jun Zheng},
  title         = {A Dynamical Lie-Algebraic Framework for Hamiltonian Engineering and Quantum Control},
  year          = {2026},
  eprint        = {2603.04916},
  archivePrefix = {arXiv},
  primaryClass  = {quant-ph},
  doi           = {10.48550/arXiv.2603.04916}
}

@article{Bethe1931,
  author  = {Bethe, H.},
  title   = {Zur Theorie der Metalle. I. Eigenwerte und Eigenfunktionen der linearen Atomkette},
  journal = {Z. Phys.},
  volume  = {71},
  pages   = {205--226},
  year    = {1931}
}

@book{Baxter1982,
  author    = {Baxter, R. J.},
  title     = {Exactly Solved Models in Statistical Mechanics},
  publisher = {Academic Press},
  address   = {London},
  year      = {1982}
}

@article{Hubbard1963,
  author  = {Hubbard, J.},
  title   = {Electron Correlations in Narrow Energy Bands},
  journal = {Proc. R. Soc. London A},
  volume  = {276},
  pages   = {238--257},
  year    = {1963}
}

@article{LiebWu1968,
  author  = {Lieb, E. H. and Wu, F. Y.},
  title   = {Absence of Mott Transition in an Exact Solution of the Short-Range, One-Band Model in One Dimension},
  journal = {Phys. Rev. Lett.},
  volume  = {20},
  pages   = {1445--1448},
  year    = {1968}
}


\clearpage
\onecolumngrid

\begin{center}
{\large\bfseries Supplemental Material for\\[4pt]
``Connectivity Controls Variational Accessibility in
Symmetry-Preserving Quantum Circuits''}
\end{center}

\vspace{0.5cm}
\twocolumngrid
\section{Single- and Double-Excitation Generators}
\label{sec:excitation_operators}

The symmetry-preserving circuits considered in this work are constructed from single- and double-excitation rotations. For the XXZ model, the single-excitation (SE) gate acting on spins \(i\) and \(j\) is
\begin{equation}
U_{\rm SE}^{(i,j)}(\theta)
=
\exp\left[
\frac{\theta}{2}
\left(
\sigma_i^{+}\sigma_j^{-}
-
\sigma_i^{-}\sigma_j^{+}
\right)
\right],
\label{eq:supp_SE}
\end{equation}
which generates a rotation within the single-excitation subspace,
\begin{equation}
|01\rangle_{ij}\leftrightarrow |10\rangle_{ij},
\end{equation}
while leaving \(|00\rangle\) and \(|11\rangle\) unchanged. This operation corresponds to coherent spin exchange and preserves the total magnetization \(S_z^{\rm tot}\). The same spin-exchange form is used for both nearest-neighbor and additional non-nearest-neighbor connections in the XXZ circuit.

For the Hubbard model, the single-excitation rotations are instead implemented as fermionic excitations,
\begin{equation}
U_{\rm FSE}^{(p,q)}(\theta)
=
\exp\left[
\frac{\theta}{2}
\left(
c_p^\dagger c_q-c_q^\dagger c_p
\right)
\right],
\label{eq:supp_FSE}
\end{equation}
where \(p\) and \(q\) denote fermionic modes belonging to the same spin species. Under the Jordan--Wigner transformation, for \(p<q\) the corresponding generator contains the parity string over all intermediate modes,
\begin{equation}
\begin{aligned}
c_p^\dagger c_q-c_q^\dagger c_p
\;\longrightarrow\;&
\frac{i}{2}\Big(
Y_p Z_{p+1}\cdots Z_{q-1}X_q
\\
&-
X_p Z_{p+1}\cdots Z_{q-1}Y_q
\Big).
\end{aligned}
\label{eq:supp_FSE_JW}
\end{equation}
up to an overall sign convention. Here \(X_p\), \(Y_p\), and \(Z_p\) denote the Pauli operators acting on qubit \(p\), and \(Z_{p+1}\cdots Z_{q-1}\) is the Jordan--Wigner parity string over the qubits lying between modes \(p\) and \(q\). This string accounts for the fermionic sign associated with the occupation parity of the intermediate modes. Thus, in the occupation basis the rotation still exchanges a fermion between modes \(p\) and \(q\), but with the fermionic sign determined by the parity of the occupied intermediate modes. The rotations are applied independently within the spin-\(\uparrow\) and spin-\(\downarrow\) sectors and therefore preserve \(N_\uparrow\) and \(N_\downarrow\) separately.

Correlated processes in the Hubbard circuit are introduced through the four-qubit double-excitation (DE) gate
\begin{equation}
U_{\rm DE}^{(i,j,k,l)}(\theta)
=
\exp\left[
\frac{\theta}{2}
\left(
\sigma_i^{+}\sigma_j^{+}\sigma_k^{-}\sigma_l^{-}
-
\sigma_i^{-}\sigma_j^{-}\sigma_k^{+}\sigma_l^{+}
\right)
\right],
\label{eq:supp_DE}
\end{equation}
which coherently couples the two configurations
\begin{equation}
|0011\rangle\leftrightarrow|1100\rangle .
\end{equation}
Depending on the assignment and ordering of the four spin orbitals, this gate realizes two physically distinct correlated processes.

For a pair of sites \(a\) and \(b\), with orbital ordering
\((a\uparrow,a\downarrow,b\uparrow,b\downarrow)\), the \emph{doublon-transfer} excitation couples
\begin{equation}
|00;\uparrow\downarrow\rangle
\leftrightarrow
|\uparrow\downarrow;00\rangle ,
\end{equation}
or, equivalently, in occupation-number notation,
\begin{equation}
|00;11\rangle\leftrightarrow|11;00\rangle .
\end{equation}

The \emph{exchange-type} double excitation instead uses the orbital ordering
\((a\uparrow,b\downarrow,a\downarrow,b\uparrow)\) and couples
\begin{equation}
|\uparrow_a,\downarrow_b\rangle
\leftrightarrow
|\downarrow_a,\uparrow_b\rangle .
\label{eq:supp_exchange}
\end{equation}
Thus, the doublon-transfer and exchange-type DE generators represent distinct correlated rearrangements of the two-particle configuration, while both preserve \(N_\uparrow\) and \(N_\downarrow\) separately. Unlike the quadratic fermionic SE rotations, these DE generators directly connect correlated many-body configurations and provide additional directions in the symmetry-preserving variational space.

\section{Dynamical Lie-algebra analysis}
\label{sec:supp_dla}
We determine the dynamical Lie algebra (DLA) generated by each circuit by iteratively evaluating commutators of the corresponding anti-Hermitian generators and retaining linearly independent directions until closure. The calculation is performed directly within each conserved symmetry sector of dimension $D$. For the XXZ model,
\begin{equation}
D=\binom{N}{N_\uparrow},
\end{equation}
whereas for the Hubbard model,
\begin{equation}
D=\binom{N}{N_\uparrow}\binom{N}{N_\downarrow}.
\end{equation}
Since the circuit generators considered here are represented by real antisymmetric matrices in the symmetry-sector basis, the maximal accessible algebra is $\mathfrak{so}(D)$, with
\begin{equation}
d_{\rm full}\equiv\dim\mathfrak{so}(D)=\frac{D(D-1)}{2}.
\end{equation}


\begin{table}[]
\caption{Dynamical Lie-algebra dimensions for the XXZ circuits in different fixed-$N_\uparrow$ symmetry sectors. $\mathcal{A}^{\rm XXZ}_{\rm NN}$ contains only NN single-excitation generators, whereas $\mathcal{A}^{\rm XXZ}_{\rm NN+1}$ contains one additional NNN single-excitation generator. Here $D=\binom{N}{N_\uparrow}$ and $d_{\rm full}=D(D-1)/2$.}
\label{tab:supp_dla_xxz}
\centering
\begin{ruledtabular}
\begin{tabular}{cccccc}
$N$ & $N_\uparrow$ & $D$ &
$d_{\rm Lie}^{\rm NN}$ &
$d_{\rm Lie}^{\rm NN+1}$ &
$d_{\rm full}$ \\
\hline
4 & 2 & 6  & 6  & 15   & 15   \\
6 & 3 & 20 & 15 & 190  & 190  \\
8 & 2 & 28 & 28 & 378  & 378  \\
8 & 3 & 56 & 28 & 1540 & 1540 \\
8 & 4 & 70 & 28 & 2415 & 2415 \\
\end{tabular}
\end{ruledtabular}
\end{table}

As shown in Table~\ref{tab:supp_dla_xxz}, the NN circuit remains restricted to
\begin{equation}
d_{\rm Lie}^{\rm NN}=\frac{N(N-1)}{2},
\end{equation}
independent of the dimension of the many-body symmetry sector. In contrast, adding a single NNN single-excitation generator yields
\begin{equation}
d_{\rm Lie}^{\rm NN+1}
=\frac{D(D-1)}{2}
=\dim\mathfrak{so}(D)
\end{equation}
for every sector examined. In particular, the $N=8$ results for $N_\uparrow=2,3,$ and $4$ demonstrate that the reconstruction happens to $S_z^{tot}\ne0$ sector too.


\begin{table}[]
\caption{Dynamical Lie-algebra dimensions for the Hubbard circuits in different fixed-$(N_\uparrow,N_\downarrow)$ sectors for $N=4$ sites. $\mathcal{A}^{\rm Hub}_{\rm NN}$ contains NN spin-conserving single excitations, $\mathcal{A}^{\rm Hub}_{\rm NN+1}$ additionally contains non-NN single excitations, and $\mathcal{A}^{\rm Hub}_{\rm NN+d}$ contains a symmetry-preserving correlated double-excitation generator. Here $D=\binom{N}{N_\uparrow}\binom{N}{N_\downarrow}$ and $d_{\rm full}=D(D-1)/2$.}
\label{tab:supp_dla_hubbard}
\centering
\begin{ruledtabular}
\begin{tabular}{ccccccc}
$N$ & $(N_\uparrow,N_\downarrow)$ & $D$ &
$d_{\rm Lie}^{\rm NN}$ &
$d_{\rm Lie}^{\rm NN+1}$ &
$d_{\rm Lie}^{\rm NN+d}$ &
$d_{\rm full}$ \\
\hline
4 & $(1,3)$ & 16 & 12 & 12 & 120 & 120 \\
4 & $(2,2)$ & 36 & 12 & 30 & 630 & 630 \\
\end{tabular}
\end{ruledtabular}
\end{table}

The Hubbard results in Table~\ref{tab:supp_dla_hubbard} reveal a qualitatively different connectivity requirement. Additional single-excitation connectivity does not generically reconstruct the full algebra: its effect is sector-dependent, leaving $d_{\rm Lie}=12$ in the $(1,3)$ sector and increasing it only to $30\ll630$ in the $(2,2)$ sector. In contrast, the addition of a correlated double-excitation generator gives
\begin{equation}
d_{\rm Lie}^{\rm NN+d}
=\frac{D(D-1)}{2}
=\dim\mathfrak{so}(D)
\end{equation}
in both examined sectors. Thus, while a single nonlocal single-excitation connection is sufficient for the XXZ circuits studied here, the Hubbard circuit requires a correlated many-body generator to reconstruct the full symmetry-sector algebra.
\begin{figure*}[t]
\centering
\includegraphics[width=\textwidth]{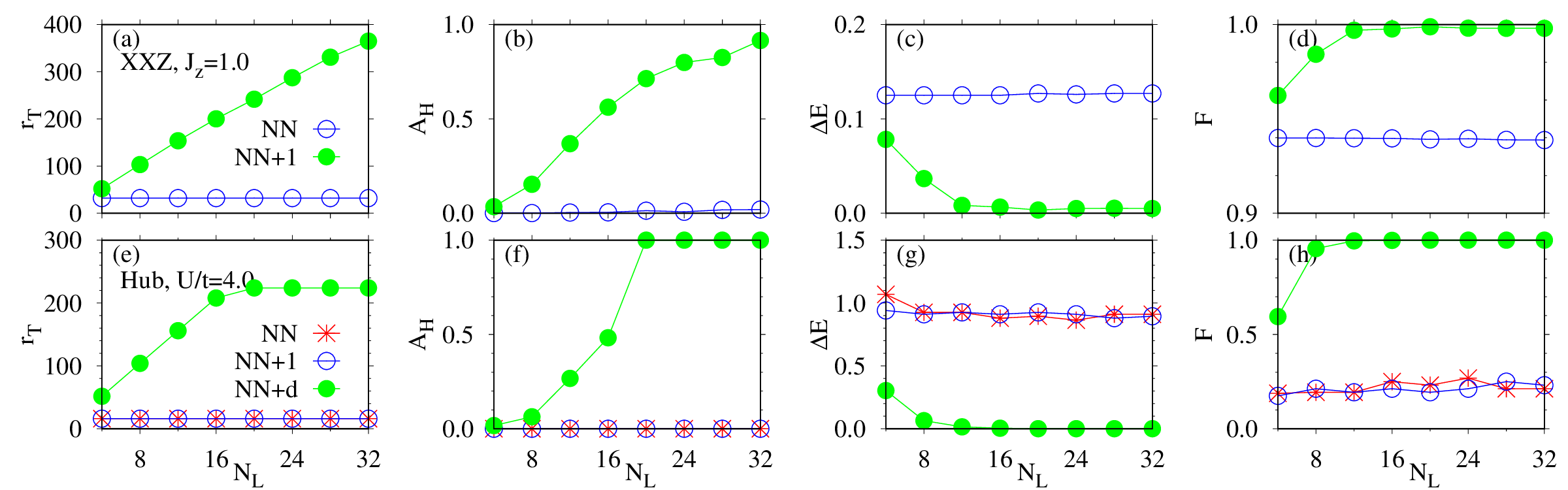}
\caption{Circuit depth ($N_L$) dependence of variational accessibility and VQE performance of various ans\"atze. For the XXZ ring (top row) with $N=12, S_z^{tot}=2$ and $J_z=1$: (a) Tangent-space rank ($r_T$), (b) Hamiltonian alignment $A_H$, (c) energy error
$\Delta E=|E_{\rm VQE}-E_0|$, and (d) ground-state fidelity
$F=|\langle\psi_0|\psi(\boldsymbol{\theta})\rangle|^2$  are shown against $N_L$ for ans\"atze $\AXXZ$ and $\AXXZOne$, indicated by empty (blue) and filled (red) circles, respectively. The corresponding results for $12-$qubit Hubbard ring with $N=6, N_\uparrow=4, N_\downarrow=2$ and $U/t=4$ are shown in (e)--(h) for ans\"atze $\AHub, \AHubOne$ and $\AHubd$ by empty (blue) circle, star (red) point and filled (green) circles, respectively.}
\label{fig:supp_xxz_sz2}
\end{figure*}

\section{VQE performance in symmetry sectors $S_z^{tot}\ne0$}
\label{sec:supp_additional_sectors}
To examine whether the connectivity-induced enhancement of variational accessibility persists beyond the symmetry sectors considered in the main text, we perform additional VQE calculations in different magnetization and particle-number sectors. We consider the $N=12$ XXZ ring in the $S_z^{\rm tot}=2$ sector and the $N=6$ Hubbard ring with $(N_\uparrow,N_\downarrow)=(4,2)$ at half filling. The same optimization procedure and variational diagnostics used in the main text are employed here.

Figure~\ref{fig:supp_xxz_sz2} shows the XXZ results for $N=12$ and $S_z^{\rm tot}=2$. The NN circuit $\mathcal{A}^{\rm XXZ}_{\rm NN}$ remains confined to a restricted tangent space, accompanied by limited accessibility of the energy-lowering Hamiltonian direction. Adding a single NNN single-excitation connection in $\mathcal{A}^{\rm XXZ}_{\rm NN+1}$ substantially enlarges the tangent-space rank $r_T$ and enhances $A_H$, resulting in a reduced energy error $\Delta E$ and increased ground-state fidelity $\mathcal F$. Thus, the improvement produced by the additional generator is not specific to the $S_z^{\rm tot}=0$ sector studied in the main text.

\begin{figure}[t]
\centering
\includegraphics[width=\columnwidth]{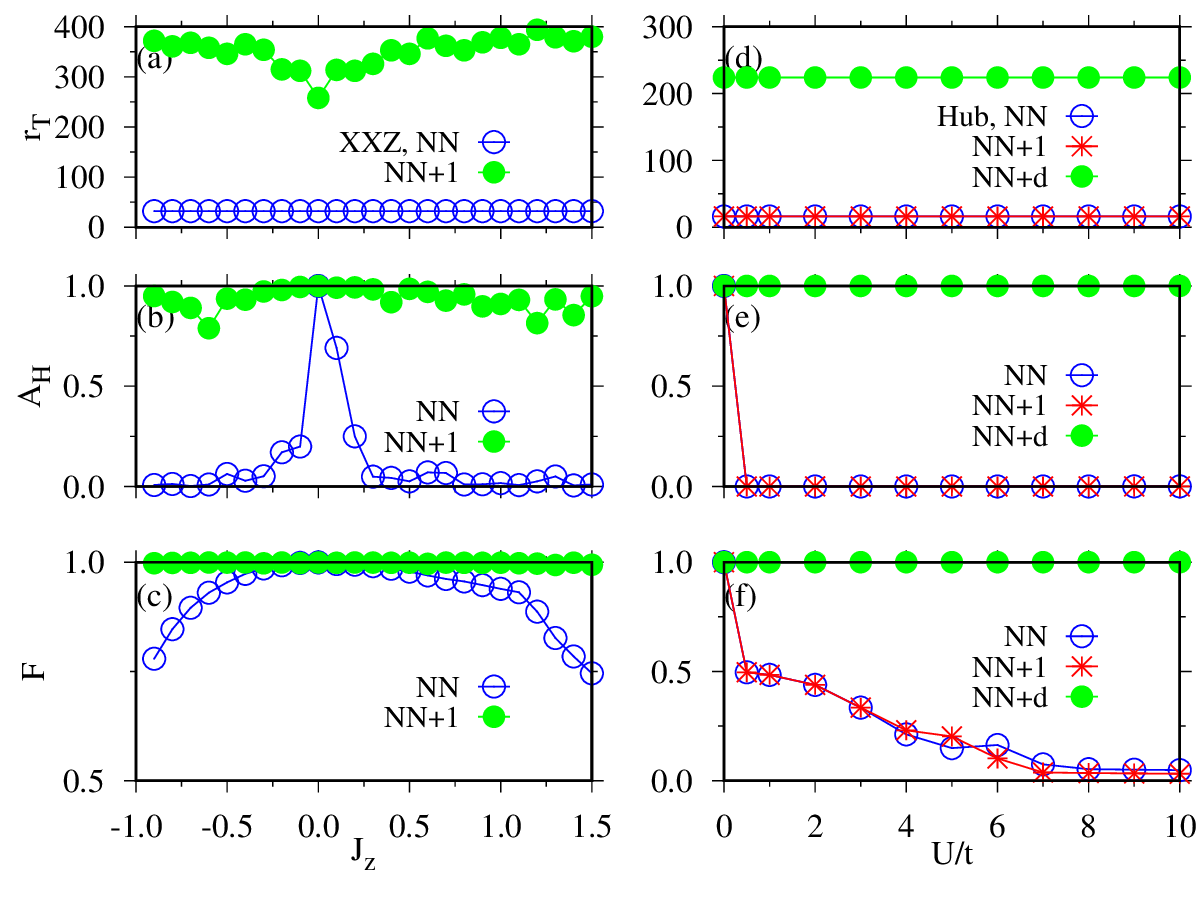}
\caption{VQE results with $N_L=32$: (a)-(c) depict tangent space rank $r_T$, Hamiltonian alignment $A_H$ and ground-state fidelity $F$   respectively against $J_z$ for spin-1/2 XXZ ring with $N=12, S^{tot}_z=2$. Empty (blue) and filled (green) circles are for ans\"atze $\AXXZ$ and $\AXXZOne$. (d)-(f) show the corresponding results against $U/t$ for $N=6, N_\uparrow=4, N_\downarrow=2$ Hubbard ring obtained for anzatz $\AHub$ (empty circle, blue), $\AHubOne$ (star point, red)  and $\AHubd$ (filled circle, green) respectively.}
\label{fig:supp_hubbard_42}
\end{figure}

A similar behavior is observed for the $N=6$ Hubbard ring in the $(N_\uparrow,N_\downarrow)=(4,2)$ sector, shown in Fig.~\ref{fig:supp_hubbard_42}. The NN single-excitation circuit $\mathcal{A}^{\rm Hub}_{\rm NN}$ remains variationally restricted, and the addition of non-NN single-excitation connectivity in $\mathcal{A}^{\rm Hub}_{\rm NN+1}$ does not remove this limitation. In contrast, supplementing the NN circuit with a symmetry-preserving correlated double excitation in $\mathcal{A}^{\rm Hub}_{\rm NN+d}$ opens additional many-body directions, leading to larger $r_T$ and $A_H$ and correspondingly improved energy and fidelity.

These additional symmetry sectors reproduce the qualitative behavior found in the main text. For XXZ, a minimal extension of the single-excitation connectivity substantially enlarges the accessible variational space, whereas for Hubbard the essential additional direction is provided by a correlated double excitation rather than by single-particle connectivity alone. The connectivity-controlled accessibility identified in the main text is therefore not tied to the particular magnetization or spin-balanced sectors used there.
\end{document}